\documentclass[journal,twoside,web]{ieeecolor}
\usepackage{generic}
\usepackage{cite}
\usepackage{amsmath,amssymb,amsfonts}
\usepackage{graphicx}
\usepackage{algorithm}
\usepackage{hyperref}
\hypersetup{hidelinks=true}
\usepackage{textcomp}
\usepackage{comment}
\usepackage{multirow}
\usepackage{booktabs}

\usepackage{array}
\usepackage{algpseudocode}

\usepackage[table]{xcolor}  
\usepackage{pifont}         

\DeclareGraphicsExtensions{.pdf,.png,.jpg}

\DeclareMathOperator*{\argmin}{argmin}

\def\BibTeX{{\rm B\kern-.05em{\sc i\kern-.025em b}\kern-.08em
    T\kern-.1667em\lower.7ex\hbox{E}\kern-.125emX}}
\begin{document}

\title{USFetal: Tools for Fetal Brain Ultrasound Compounding}

\author{Mohammad Khateri, Morteza Ghahremani, Sergio Valencia, Camilo Jaimes, \\ Alejandra Sierra, Jussi Tohka, P. Ellen Grant, Davood Karimi, Senior, IEEE
\thanks{This work was supported in part by the Research Council of Finland under Grants 323385 and 358944; by the Flagship of Advanced Mathematics for Sensing Imaging and Modelling (Grant \#358944); by the Finnish Cultural Foundation; by the Saastamoinen Foundation; by the KAUTE Foundation; by the Finnish Foundation for Technology Promotion; and by NIH grants R01NS128281 and R01HD110772. The content of this publication is solely the responsibility of the authors and does not necessarily represent the official views of the NIH.}
\thanks{
Mohammad Khateri, Alejandra Sierra, and Jussi Tohka are with the A. I. Virtanen Institute for Molecular Sciences, Faculty of Health Sciences, University of Eastern Finland, 70210 Kuopio, Finland (e-mail: \{mohammad.khateri, alejandra.sierra, jussi.tohka\}@uef.fi).
Morteza Ghahremani is with the Munich Center for Machine Learning (MCML) at the Technical University of Munich, 80333 München, Germany (e-mail: morteza.ghahremani@tum.de).
Sergio Valencia  and Camilo Jaimes are with the Department of Radiology, Massachusetts General Hospital, Boston, MA 02114, USA (e-mail: svalenciavasquez@mgh.harvard.edu; cjaimescobos@mgb.org).
P Ellen Grant and Davood Karimi are with the Harvard Medical School and Boston Children’s Hospital, Boston, MA 02115, USA (e-mail: \{ellen.grant, davood.karimi\}@childrens.harvard.edu).
}
} 
\maketitle

\begin{abstract}
Ultrasound offers a safe, cost-effective, and widely accessible technology for fetal brain imaging, making it especially suitable for routine clinical use. However, it suffers from view-dependent artifacts, operator variability, and a limited field of view, which make interpretation and quantitative evaluation challenging. Ultrasound compounding aims to overcome these limitations by integrating complementary information from multiple 3D acquisitions into a single, coherent volumetric representation.
This work provides four main contributions: 
(1) We present the first systematic categorization of computational strategies for fetal brain ultrasound compounding, including both classical techniques and modern learning-based frameworks.
(2) We implement and compare representative methods across four key categories—multi-scale, transformation-based, variational, and deep learning approaches—emphasizing their core principles and practical advantages.
(3) Motivated by the lack of full-view, artifact-free ground truth required for supervised learning, we focus on unsupervised and self-supervised strategies and introduce two new deep learning–based approaches: a self-supervised compounding framework and an adaptation of unsupervised deep plug-and-play priors for compounding. 
(4) We conduct a comprehensive evaluation on ten multi-view fetal brain ultrasound datasets, using both expert radiologist scoring and standard quantitative image-quality metrics.
We also release the \textbf{USFetal Compounding Toolbox}, publicly available to support benchmarking and future research:
\textcolor{blue}{\url{https://github.com/mkhateri/USFetal-Compounding-Tools}}.
\end{abstract}


\section{Introduction}
\label{sec:introduction}

\subsection{Background and motivation}
\label{sec:background}

\IEEEPARstart{T}{he} fetal period represents a critical stage in neurodevelopment \cite{bayer2005human}. A complex set of intricate and tightly choreographed processes take place in this period. Starting with a simple neural tube, these processes form different brain structures and their inter-connections within a short time period. The fetal brain is vulnerable to neurological disorders, prone to malformations, and sensitive to adverse environmental conditions. Brain disorders arising during the fetal development can profoundly impact the maturation of cognitive and motor functions and result in lifelong neurological impairment or disabilities \cite{donofrio2011impact, jakab2015disrupted}.

Medical imaging technologies, such as ultrasound, are indispensable for non-invasive monitoring and assessment of brain development during this critical period. Fetal ultrasound is widely used for brain imaging due to its high spatial resolution, safety, and cost-effectiveness compared to magnetic resonance imaging (MRI). It holds the potential to enhance early diagnosis, optimize fetal interventions, and mitigate the long-term impact of neurological disorders.

Currently, fetal brain ultrasound relies primarily on 2D imaging, which has limited capability for visualizing and quantitatively assessing the brain. Moreover, 2D ultrasound is susceptible to artifacts and high variability in acquisition and interpretation, which depend on the operator. The limited field of view in 2D ultrasound restricts morphometric measurements to a few key parameters, such as biparietal diameter, head circumference, and lateral ventricle width. Volumetric (3D) ultrasound offers significantly improved visualization and assessment of various brain structures, enhancing the detection of abnormalities \cite{duckelmann2010three, gonccalves2016three}, such as ventriculomegaly and cortical malformations. Despite these advantages, 3D ultrasound remains operator-dependent, requires specialized expertise for accurate interpretation, and is susceptible to view-dependent artifacts. Advanced image analysis tools are urgently needed to fully harness the potential of 3D fetal brain sonography.

The quality of fetal brain ultrasound images is compromised by factors such as a limited field of view, variations in acoustic impedance of different tissues, and fetal and probe motion. As a result, each probe orientation provides a limited view of the brain, enhancing the visualization of certain key anatomical regions—such as the ventricles, midline structures, or posterior fossa—while other regions remain outside the field of view or appear poorly defined \cite{ortiz2012ultrasound}. Reliable visualization of the relevant intracranial anatomy therefore requires combining multiple views acquired from different probe orientations into a single 3D image. Ideally, the resulting composite image will clearly depict these structures with an enhanced signal-to-noise ratio (SNR) and reduced artifacts. In the ultrasound imaging literature, this process is known as \emph{compounding} \cite{krucker20003d} and is closely related to tasks such as image fusion \cite{karim2023current}, super-resolution \cite{wang2020deep}, and image reconstruction \cite{ben2021deep}. Compounding is a crucial first step in 3D fetal brain ultrasound imaging to enable its full potential for assessing clinically relevant anatomical features and structural abnormalities.

Prior works on fetal brain ultrasound compounding have been mostly based on spatial alignment of different views followed by fusion via simple voxel-wise operations such as maximum pooling or averaging \cite{wilhjelm2004visual, behar2003new, leotta1999three}. Overall, images computed with these approaches often offer improved image quality compared to individual views. Nonetheless, these approaches fail to optimally exploit the information in different views. There have been no systematic efforts to explore the potential of advanced computational imaging methods for fetal brain ultrasound compounding. Deep learning methods, for example, have shown great potential to address related problems in computer vision. There is an urgent need for a carefully designed and systematic exploration of methods for fetal brain ultrasound compounding.

The goal of this work is to address the limitations of current fetal brain ultrasound technology by leveraging modern computational imaging techniques for systematic exploration and benchmarking of compounding methods. Based on best-performing methodologies in computational imaging, we suggest that four classes of techniques can be considered for fetal brain ultrasound compounding: multi-scale, transformation-based, variational, and deep learning methods. We explain the design principles, strengths, and weaknesses of each class. Moreover, we implement a representative method from each class and critically evaluate them on real-world 3D ultrasound data acquired from fetal brains in utero. The results of this study, together with the method implementations that we publicly release, can greatly benefit this important field and help substantially expand the potential of 3D ultrasound for assessing normal and abnormal brain development in utero.

\subsection{Related works}
\label{sec:related_works}

Enhancing image quality through the compounding of 3D ultrasound acquisitions has been explored previously by a few studies \cite{robinson1981computer, krucker20003d}. The conventional approach is to obtain uncorrelated views of the anatomy of interest, spatially align them using off-the-shelf registration techniques, and fuse them to produce a final image. Different approaches to registration, including rigid and non-rigid methods, have been attempted \cite{krucker20003d}. Likewise, a range of fusion strategies has been explored, including voxel-wise operations such as the arithmetic/geometric mean, median, and maximum \cite{wilhjelm2004visual, behar2003new, krucker20003d, leotta1999three}. More sophisticated fusion methods weigh the contribution of different views based on local structures using frequency analysis \cite{grau2005adaptive}, wavelets \cite{rajpoot2009multiview}, or image pyramids \cite{wright2019complete}. It has been argued that Laplacian and Gaussian pyramids can preserve genuine image features while suppressing noise and artifacts. Other methods, such as directed acyclic graphs \cite{hung2021ultrasound} and B-splines \cite{zimmer2018multi} may also be used to determine the confidence of different regions in each view to achieve a more informed fusion.

More recently, limited attempts have been made to use deep learning techniques for image compounding in fetal ultrasound. One study used spatial transformer networks to register different views, which were then fused using saliency-weighted averaging with Laplacian pyramids \cite{wright2019complete}. Other studies have employed deep learning models such as convolutional neural networks to perform brain segmentation and registration operations involved in compounding \cite{perez2020deep, wright2023fast}. However, to date, no published studies have systematically explored the potential of classical and state-of-the-art computational imaging methods for fetal brain ultrasound compounding.

\section{Materials and Methods}

\subsection{Problem setting}

\begin{figure}[ht]
    \centering
    \includegraphics[width=1.0\linewidth]{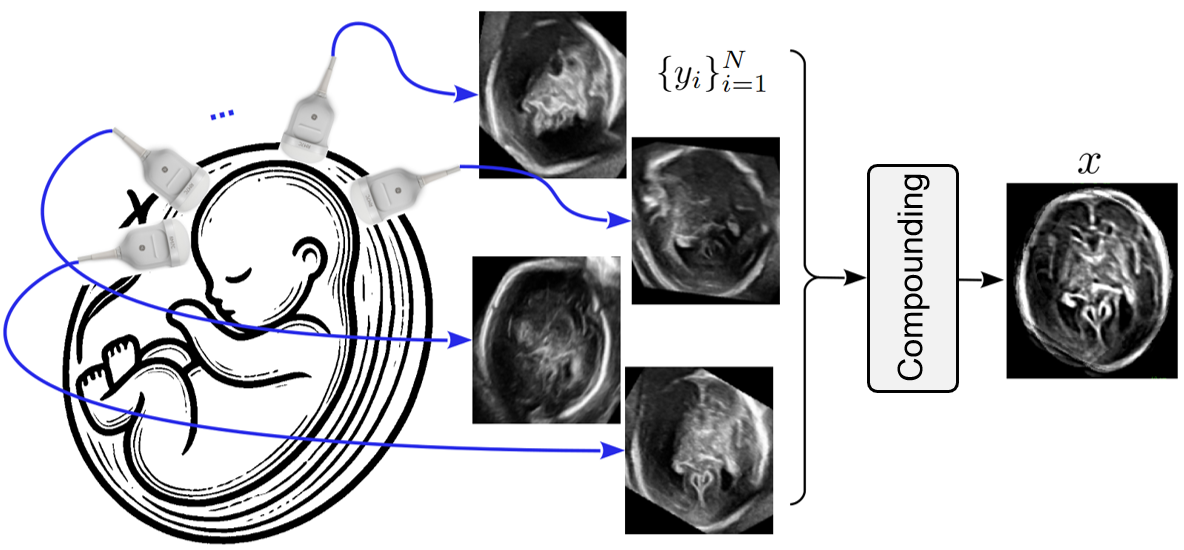}
    \caption{Schematic illustration of multi-view fetal brain ultrasound acquisition and compounding. Each acquired 3D volume, ${y}_i$, captures a partial view of the brain, where the varying probe orientation determines the viewing angle and view-dependent artifacts. The views are typically not aligned and differ significantly in appearance. Compounding aims to integrate these complementary, imperfect views into a coherent, high-quality representation of the whole brain.}
\label{fig:acquisition}
\end{figure}

In clinical settings, 3D fetal brain ultrasound volumes are acquired using multiple probe orientations, each yielding a restricted field of view and exhibiting orientation-dependent artifacts, including shadowing, reverberation, refraction, and speckle noise \cite{ortiz2012ultrasound}. Let \({y}_i \in \mathbb{R}^{D \times H \times W}\) denote the \(i\)-th acquired view and \(x \in \mathbb{R}^{D \times H \times W}\) be the desired full-view image. We can model the observation process as,
\begin{equation}
{y}_i = \mathcal{H}_{\delta_i}(x), \quad i = 1,\dots,N,
\end{equation}
where \(\mathcal{H}_{\delta_i}\) represents the imaging system, parameterized by \(\delta_i\) to incorporate the effects of the ultrasound imaging (including spatial transformations, noise, and directionality) associated with the \(i\)-th view. Fig~\ref{fig:acquisition} illustrates the multi-view acquisition process. The goal of ultrasound compounding is to integrate the complementary information available in \(\{{y}_i\}_{i=1}^N\) into one high-quality image. However, reconstructing the full-view image $x$ from these partial and noisy observations is an ill-posed inverse problem.

\subsection{Compounding methods}

The computational imaging literature offers a wide range of solutions that can be harnessed to address the challenges of ultrasound compounding. These methods appear under such applications as image fusion \cite{ singh2023review}, super-resolution \cite{khateri2025mri}, and reconstruction \cite{qiu2023medical}. Here, we present the first systematic study of applicable approaches for ultrasound compounding, categorizing these techniques from traditional methods to modern deep learning–based frameworks. We particularly emphasize unsupervised and self-supervised approaches that operate without full-view ground-truth data, the acquisition of which is infeasible in clinical ultrasound imaging.

\subsubsection{Multi-scale Approaches}

Multi-scale fusion offers an unsupervised framework for integrating complementary information across images at different resolutions. It is particularly well-suited to ultrasound imaging, where anatomical structures exhibit considerable variability in both scale and visual appearance. These methods typically decompose each view into hierarchical representations to isolate structural features across distinct spatial frequencies or scales. Fusion is then performed at each scale, and the results are integrated to yield the final compounded volume.

Let each view $y_i$ be decomposed into $L$ scales via an operator $\mathcal{T}$:
\begin{equation}
\mathcal{T}(y_i) = \{\, y_i^{(\ell)} \,\}_{\ell=1}^{L},
\end{equation}
where $y_i^{(\ell)}$ denotes the $\ell$-th scale representation of view $i$. Fusion is performed at each scale using a rule $\mathcal{F}^{(\ell)}$ (e.g., averaging, weighted sum, or max selection), and the fused scales are reconstructed by the inverse transform:
\begin{align}
z^{(\ell)} &= \mathcal{F}^{(\ell)}\!\big(y_1^{(\ell)},\,\dots,\,y_N^{(\ell)}\big), \qquad \ell=1,\dots,L, \\
x^{*} &= \mathcal{T}^{-1}\!\big(\{\, z^{(\ell)} \,\}_{\ell=1}^{L}\big).
\end{align}
where $x^{*}$ denotes the final compounded volume.

This decomposition enables fine-grained control over the fusion process, allowing selective emphasis or suppression of features such as edges, textures, and local contrast at different scales. Representative techniques include Gaussian and Laplacian pyramids~\cite{li2018gaussian,jiang2023lightweight}, Difference-of-Gaussians (DoG)~\cite{jie2023medical}, and wavelet-based decompositions~\cite{li1995multisensor}.

\subsubsection{Transformation-based Approaches}
Transformation-based compounding re-expresses \emph{multi-view} data in a domain where salient structures become more separable. Techniques such as principal component analysis (PCA) and independent component analysis (ICA) project the input data onto orthogonal or statistically independent bases, enabling compact and effective fusion while reducing redundancy. In ultrasound compounding, the $N$ registered views can be fused \emph{jointly}: at each spatial location $\mathbf{p}$, the view vector $\mathbf{y}(\mathbf{p}) = [y_1(\mathbf{p}), \dots, y_N(\mathbf{p})]^\top$ is first transformed into component maps via $\mathcal{T}$, then fused in the component domain, and finally reconstructed:
\begin{align}
\mathbf{s}(\mathbf{p}) &= \mathcal{T}\big(\mathbf{y}(\mathbf{p})\big), \\
\hat{\mathbf{s}}(\mathbf{p}) &= \mathcal{F}\big(\mathbf{s}(\mathbf{p})\big), \\
x^{*}(\mathbf{p}) &= \mathcal{T}^{-1}\big(\hat{\mathbf{s}}(\mathbf{p})\big).
\end{align}
Here, $\mathcal{F}$ denotes a component-wise fusion rule—either fixed or adaptive—that determines the relative emphasis across components. This joint formulation captures dominant cross-view correlations while effectively suppressing redundancy and noise. Representative transformations such as PCA~\cite{kumar2006pca, he2010multimodal} and ICA~\cite{calhoun2008ica, adali2015multimodal} have been extensively employed in image fusion applications.

\subsubsection{Variational Fusion Framework}
Variational approaches offer a principled unsupervised framework for image fusion by casting the task as an energy minimization problem \cite{wang2023novel, khateri2020variational}. Specifically, the fused volume \(x\) is estimated as the solution of the following optimization problem:
\begin{equation}
x^* = \arg \min_x \left\{ \mathcal{D}(x; \{{y}_i\}_{i=1}^N) + \mu\, \mathcal{R}(x) \right\},
\end{equation}
where \(\mathcal{D}(\cdot)\) is a fidelity term enforcing consistency between the fused image and the input views \(\{{y}_i\}_{i=1}^N\), and \(\mathcal{R}(\cdot)\) is a regularization term promoting prior knowledge. A standard choice for the fidelity term is the voxel-wise \(\ell_p\) deviation:
\begin{equation}
\mathcal{D}_{\text{voxel}}(x; \{{y}_i\}_{i=1}^N) = \sum_i \|x - {y}_i\|_p^p,
\end{equation}
which penalizes intensity mismatch at voxel level. While simple and effective, this formulation ignores structural consistency.
To capture richer correspondences across views, we can consider the fidelity terms in feature domains using a differentiable feature operator \(\phi(\cdot)\):
\begin{equation}
\mathcal{D}_{\phi}(x; \{{y}_i\}_{i=1}^N) = \sum_i \|\phi(x) - \phi({y}_i)\|_q^q,
\end{equation}
where \(\phi(\cdot)\) encodes domain-specific image features (e.g., gradients, band-pass filters), leading to the extended formulation:

\begin{align}
\label{variational_formula}
x^* = \arg \min_x \Bigg\{ 
&\underbrace{\sum_i \|x - y_i\|_p^p}_{\textbf{Image Fidelity}} 
+ \lambda\, \underbrace{\sum_i \|\phi(x) - \phi(y_i)\|_q^q}_{\textbf{Feature Consistency}} \notag\\
&+ \mu\, \underbrace{\mathcal{R}(x)}_{\textbf{Regularization}} 
\Bigg\}
\end{align}
where \(\lambda, \mu > 0\) control the relative influence of the fidelity and regularization terms, and \(p, q \in \{1, 2\}\) specify the norm types used in the respective components. Regularization incorporates prior knowledge about the data, typically through total variation~\cite{rudin1992nonlinear}, sparsity~\cite{jin2017sparsity}, or non-local priors~\cite{zha2023learning}. The objective function (\ref{variational_formula}) can be solved through a wide range of established optimization methods \cite{boyd2004convex, beck2017first}.

\begin{figure*}[h]
    \centering
    \includegraphics[width=1.0\linewidth]{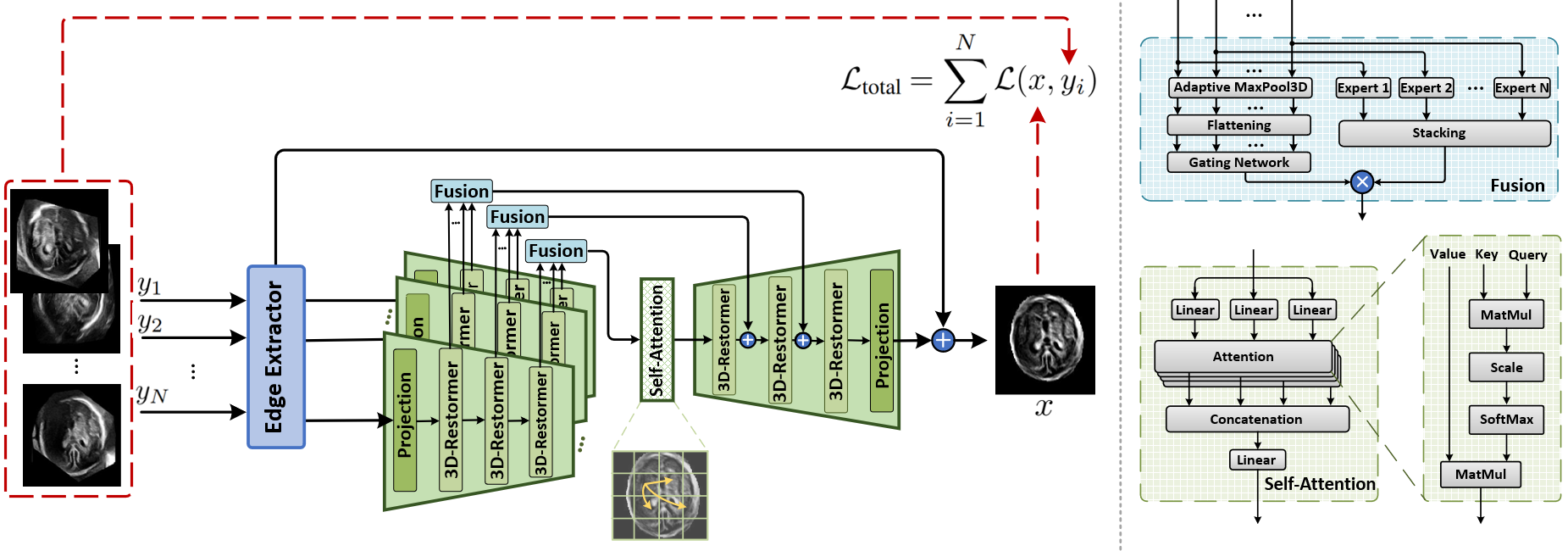}
    \caption{Schematic of the proposed self-supervised compounding architecture. The input volumes $\{y_i\}_{i=1}^N$ are first processed by an edge extractor module, which captures anatomical boundaries and forwards them via skip connections. Each view is then projected into a shared feature space and encoded across multiple spatial scales. The encoded features are fused and decoded to form a preliminary reconstruction, which is subsequently refined by integrating the skipped edge information. The resulting compounded volume is compared to the input views using a self-supervised loss that enforces both voxel-wise and structural consistency, enabling end-to-end optimization without the need for ground-truth supervision.}
    \label{fig:USfetal_diag}
\end{figure*}

\subsubsection{Deep Learning Approaches}

Learning-based approaches aim to learn a parameterized mapping \( f_\theta(\cdot) \) that combines multiple input views into a single coherent representation. Given a multi-view input \(Y = \{y_i\}_{i=1}^{N}\), the network produces a compound output volume:
\begin{equation}
x^* = f_\theta({Y}),
\end{equation}
where \(f_\theta: \mathbb{R}^{N \times D \times H \times W} \rightarrow \mathbb{R}^{D \times H \times W}\) maps the \(N\) input views to a single volumetric estimate. The parameters \(\theta\) are optimized by minimizing the expected loss over a dataset of \(M\) pairs of training samples $\{Y^{(j)}, x^{(j)}\}_{j=1}^{M}$:
\begin{equation}
\theta^* = \argmin_{\theta} \; \frac{1}{M} \sum_{j=1}^{M} \mathcal{L}(f_\theta({Y}^{(j)}), x^{(j)}),
\end{equation}
where ${Y}^{(j)}$ denotes the multi-view input for the \(j\)-th training sample, \(x^{(j)}\) is the compounded target volume, and \(\mathcal{L}(\cdot)\) is a loss function measuring the discrepancy between the network output and the target.

In the context of ultrasound imaging, supervised learning is generally infeasible due to the absence of full-view, artifact-free ground-truth data. To address this limitation, we investigate two complementary strategies for fetal ultrasound compounding: a self-supervised learning framework and a training-free deep plug-and-play (PnP) approach adapted to this task.

\vspace{0.5em}
\noindent\textbf{Self-Supervised Learning:}
Self-supervised methods exploit the inherent consistency across multiple views of a scan to enable learning without reliance on ground-truth views. The network is optimized by minimizing both voxel-wise and structural discrepancies between the output and the input views. This paradigm is particularly well-suited for ultrasound imaging, where inter-view redundancy serves as the primary supervisory signal in the absence of full-view ground truth.
Given a multi-view input \({Y}^{(j)}\), the network \(f_\theta(\cdot)\) produces a compound output \(x^{(j)*} = f_\theta({Y}^{(j)})\). The total self-supervised loss for the \(j\)-th sample is defined as:
\begin{align}
\mathcal{L}_{\text{self}} & (x^{(j)*}, {Y}^{(j)}) 
= \notag \\
&\underbrace{\sum_{i=1}^{N_j} \|x^{(j)*} - Y_i^{(j)}\|_p^p}_{\textbf{Voxel-wise Consistency}} 
+ \lambda \underbrace{\sum_{i=1}^{N_j} \|\phi(x^{(j)*}) - \phi(Y_i^{(j)})\|_q^q}_{\textbf{Feature-level Consistency}}, \label{eq:self-supervised-loss}
\end{align}

\noindent
where \(Y_i^{(j)}\) denotes the \(i\)-th view from the \(j\)-th sample, $N_j$ is the number of views in \(j\)-th sample, and \(\phi(\cdot)\) is a differentiable feature extractor, such as the DoG operator, used to capture structural patterns across a range of spatial frequencies. The first term encourages voxel-wise fidelity to the input views, while the second term promotes structural alignment. The network is trained by minimizing the expected self-supervised loss across all training samples:
\begin{equation}
\theta^* = \arg\min_{\theta} \; \frac{1}{M} \sum_{j=1}^{M} \mathcal{L}_{\text{self}}(x^{(j)*}, {Y}^{(j)}).
\end{equation}

\textit{Scan-specific self-supervised learning:} When only a single scan is available (i.e., \(M = 1\)), training becomes scan-specific, relying solely on intra-scan redundancy.

\subsubsection*{Network Architecture}
The proposed self-supervised network for compounding multi-view volumetric ultrasound images, as shown in Fig.~\ref{fig:USfetal_diag}, comprises two main components: (i) an \textit{edge extractor} and (ii) a \textit{feature fusion module}.

\paragraph{Edge Extractor}  
This module, denoted by $\mathcal{M}_{FE}$, focuses on extracting mid-frequency anatomical boundaries using a DoG operator. These features capture key structural regions within the fetal brain. The extracted edge information is skipped directly to the output, helping preserve salient anatomy while allowing the network to concentrate on learning complementary details across views:
\begin{equation}
    x^{(j)} = \mathcal{M}_{FE}(Y^{(j)}).
\end{equation}

\paragraph{Feature Fusion}  
Each input volume is projected into a shared feature space and processed hierarchically using a multi-scale U-shaped encoder-decoder architecture. The encoder is shared across all views to extract consistent representations. At each scale, a dedicated fusion block integrates features through two parallel branches.

In the first branch, feature maps are passed through the restormer, a set of expert modules, each implemented as a 3D convolutional layer with a ReLU activation. These modules capture diverse local patterns and are designed to specialize in different spatial or anatomical features. Their outputs are stacked to form a rich representation space.

In parallel, the second branch performs 3D adaptive max pooling on the input features, followed by flattening. The resulting vector is passed through a gating network, implemented as a fully connected layer followed by a softmax activation, which generates normalized fusion weights corresponding to each view. These weights are used to modulate the stacked expert outputs via element-wise multiplication, allowing the network to adaptively emphasize relevant features based on input context.

The fused features are further refined using a self-attention block, which captures long-range dependencies and enhances spatial coherence. The resulting feature maps are decoded through the hierarchical decoder, projected back into the image domain, and combined with the skipped edge features to produce the final compounded output:
\begin{equation}
    x^{(j)} = \mathcal{M}_{FF}(x^{(j)}).
\end{equation}

\paragraph{Training with Self-Supervision}  
The final output volume is compared with the input views using a self-supervised loss that enforces both voxel-level fidelity and structural consistency in a learned feature space. The loss is backpropagated through the network, and the model parameters are updated accordingly during training.

\textbf{Fusion by Mixture of Experts:}
The restormer block is built upon a Mixture of Experts (MoE) framework. Given that a shared encoder provides a feature pyramid comprising \(K\) feature levels over \(N\) views, denoted as,
\[
\mathbf{F}_k = \{\mathbf{F}_1^k, \ldots, \mathbf{F}_N^k\}, \quad k=1,\dots,K,
\]
we define \(K\) fusion blocks, each mapping the feature maps from \(N\) views into a unified representation:
\[
\mathcal{F}^k : \mathbb{R}^{N \times C_k \times D_k \times H_k \times W_k} \rightarrow \mathbb{R}^{C_k \times D_k \times H_k \times W_k}.
\]
Since the contribution of each view depends on the amount of information it contains, we employ an MoE mechanism to efficiently compute each view’s contribution to the fused feature map. To this end, we first define a gating network \(G\) that determines the contribution weights of each view. The gating network applies adaptive max pooling over the spatial dimensions, then flattens the resulting features along with the channel features into a vector \(\mathbf{g}\). This vector is passed through a fully connected layer to produce the set of coefficients \(\{G\}\), where:
\[
G : \mathbb{R}^{N \times C_k \times D_k \times H_k \times W_k} \rightarrow \mathbb{R}^{N},
\]
and the weights are computed as:
\begin{equation}
G_n(\mathbf{g}) = \frac{\exp(\mathbf{w}_n^\top \mathbf{g} + b_n)}{\sum_{j=1}^N \exp(\mathbf{w}_j^\top \mathbf{g} + b_j)},
\end{equation}
where the contributions satisfy:
\[
\sum_{n=1}^N G_n(\mathbf{g}) = 1, \quad \text{and} \quad G_n(\mathbf{g}) \ge 0.
\]
$\mathbf{w}_kn$ and $b_n$ are the learnable parameters of the gating network for the $n$-th expert. 

\noindent \textbf{Experts:} Each expert aims to extract high-level features from each encoded input feature, i.e.,
\[
\mathbf{F}_n^k \rightarrow \mathbf{E}_n^k,
\]
where the fused feature embedding \(\mathbf{A}_k\) is computed as a weighted sum of the outputs from the \(N\) experts' features:
\begin{equation}
\mathbf{A}_k = \sum_{n=1}^N G_n(\mathbf{g}) \cdot \mathbf{E}_n^k.
\end{equation}

Inspired by \cite{zamir2022restormer,zhao2024equivariant}, we utilize a dual feature extraction framework composed of a global and a local component, forming a CNN-Transformer hybrid architecture. The global feature extractor is a vision Transformer that leverages self-attention mechanisms to explore and model long-range dependencies and interactions between voxels across the entire volume. This allows the model to understand contextual relationships that span large spatial extents. In contrast, the local feature extractor is a CNN-based residual block that focuses on capturing fine-grained local variations and details, which are critical for precise localization and subtle feature recognition. By combining these two complementary modules, our approach aims to effectively leverage both global context and local details, leading to improved performance in complex medical imaging tasks.

\begin{algorithm}[t]
\caption{Plug-and-Play Ultrasound Compounding}
\label{alg:pnp-fusion}
\begin{algorithmic}[1]
\State \textbf{Input:} Pre-aligned 3D volumes $\{y_i\}_{i=1}^N$; denoiser $\mathcal{D}_\sigma$; directions $\mathcal{P} = \{\text{axial, coronal, sagittal}\}$; step size $\eta > 0$; iterations $T$
\State \textbf{Output:} Compounded volume $x^*$
\vspace{0.5em}
\State Compute initial average: $\bar{y} \gets \frac{1}{N} \sum_{i=1}^N y_i$
\For{each direction $p \in \mathcal{P}$}
    \State Initialize directional volume: $\hat{x}_p \gets 0$
    \For{each slice index $k$ along direction $p$}
        \State Initialize: $s_k^{(0)} \gets$ $k$-th slice of $\bar{y}$
        \For{$t = 0$ to $T{-}1$}
            \State \textbf{Fidelity step:} $z_k^{(t)} \gets (1 - \eta)\, s_k^{(t)} + \eta\, s_k^{(0)}$
            \State \textbf{Denoising step:} $s_k^{(t+1)} \gets \mathcal{D}_\sigma(z_k^{(t)})$
        \EndFor
        \State Insert final slice: $\hat{x}_p[k] \gets s_k^{(T)}$
    \EndFor
\EndFor
\State \textbf{Return:} $x^* \gets \frac{1}{|\mathcal{P}|} \sum_{p \in \mathcal{P}} \hat{x}_p$
\end{algorithmic}
\end{algorithm}

\vspace{0.5em}
\noindent\textbf{Training-free (Unsupervised) Methods.}
Plug-and-Play (PnP) methods provide a modular framework for solving inverse imaging problems without task-specific training. Originally introduced for image restoration, PnP decouples data fidelity and prior modeling by embedding a pretrained denoiser as an implicit regularizer within an iterative optimization scheme~\cite{romano2017little,zhang2021plug}. The general objective is formulated as:
\begin{equation}
x^* = \arg\min_{x} \; \mathcal{H}(x, y) + \lambda\, \mathcal{R}(x),
\end{equation}
where \(\mathcal{H}(x, y)\) enforces fidelity to the observed data \(y\), and \(\mathcal{R}(x)\) encodes prior knowledge. In the PnP framework, \(\mathcal{R}(x)\) is not defined explicitly but is approximated via a denoising operator \(\mathcal{D}_\sigma\), resulting in the iterative scheme:
\begin{align}
\textbf{Fidelity step:} \quad & z^{(t)} = x^{(t)} - \eta\, \nabla_x \mathcal{H}(x^{(t)}, y), \\
\textbf{Denoising step:} \quad & x^{(t+1)} = \mathcal{D}_\sigma(z^{(t)}),
\end{align}
with step size \(\eta > 0\) and iteration index \(t\).

\medskip
\noindent
To adapt PnP for multi-view ultrasound compounding, we treat the task as a volumetric image restoration problem. Let \(\{y_i\}_{i=1}^N\) denote a set of pre-aligned 3D ultrasound volumes. We first compute a coarse initialization via voxel-wise averaging $\bar{y} = \frac{1}{N} \sum_{i=1}^N y_i$, which serves as a degraded observation of the latent compounded volume \(x\). Although ultrasound noise is typically non-Gaussian and signal-dependent, this averaging suppresses structured artifacts and yields residual noise that can be reasonably approximated as Gaussian, thereby justifying the use of Gaussian-trained denoisers in the PnP framework.

\medskip
\noindent
We then apply an iterative refinement procedure, alternating between fidelity and denoising steps. A pretrained 2D denoiser (e.g., DRUNet~\cite{zhang2021plug}) is applied slice-wise along the axial, coronal, and sagittal planes. The final compounded volume is obtained by averaging the outputs across all directions, preserving both local detail and global anatomical structure. The complete procedure is described in Algorithm~\ref{alg:pnp-fusion}.

\subsection{Quality Assessment Metrics}

\subsubsection{Quantitative Metrics}
In the absence of artifact-free ground truth, evaluating the quality of compounded fetal brain ultrasound volumes is inherently nontrivial. To address this, we utilize several well-established quantitative metrics that measure how effectively the compounded image $x$ integrates and preserves information from the input volumes $\{{y}_i\}_{i=1}^N$. Each metric is described below using a general pair of images $A$ and $B$ (e.g., $x$ and ${y}_i$), followed by how it is applied in the multi-view fusion context.

\vspace{1ex}
\textit{Mutual Information (MI) \cite{mackay2003information, pedregosa2011scikit}:}  
MI quantifies the amount of shared statistical information between two images. For images $A$ and $B$, it is defined as:
\begin{equation}
\text{MI}(A, B) = \sum_{a \in \mathcal{A}} \sum_{b \in \mathcal{B}} p_{A,B}(a,b) \log \left( \frac{p_{A,B}(a,b)}{p_A(a)\,p_B(b)} \right),
\end{equation}
where $p_{A,B}$ is the joint distribution of intensities, and $p_A$, $p_B$ are the marginal distributions.

\vspace{1ex}
\textit{Structural Similarity Index (SSIM)\cite{wang2004image}:}  
SSIM assesses perceptual similarity by comparing structure, luminance, and contrast. It is defined as:
\begin{equation}
\text{SSIM}(A, B) = \frac{(2\mu_A \mu_B + C_1)(2\sigma_{AB} + C_2)}{(\mu_A^2 + \mu_B^2 + C_1)(\sigma_A^2 + \sigma_B^2 + C_2)},
\end{equation}

where $\mu$, $\sigma^2$, and $\sigma_{AB}$ denote the mean, variance, and covariance of $A$ and $B$, respectively.

\vspace{1ex}
\textit{Entropy (H) \cite{shannon1948mathematical}:}  
Entropy measures the overall information richness of an image $A$ as:
\begin{equation}
H(A) = -\sum_{a \in \mathcal{A}} p_A(a) \log p_A(a),
\end{equation}
where $p_A$ is the intensity distribution in $A$.

\vspace{1ex}
\textit{Correlation Coefficient (CC):}  
CC captures the linear relationship between two images. It is given by:
\begin{equation}
\text{CC}(A, B) = \frac{\text{cov}(A, B)}{\sigma_A \sigma_B},
\end{equation}
where $\text{cov}(A, B)$ is the covariance, and $\sigma$ is the standard deviation. 

\vspace{1ex}
\noindent\textbf{Metric Averaging Across Views:}  
For MI, SSIM, and CC, we compute an average over all $N$ input views to assess global consistency:
\begin{equation}
\text{Metric}_{\text{avg}}(x) = \frac{1}{N} \sum_{i=1}^N \text{Metric}(x, \mathbf{y}_i).
\end{equation}

\vspace{1ex}
\begin{itemize}
    \item \textbf{MI} assesses shared information between the fused output and each input.
    \item \textbf{SSIM} evaluates perceptual similarity and structure preservation.
    \item \textbf{Entropy} reflects the information richness of the final fused volume.
    \item \textbf{CC} quantifies global linear similarity.
\end{itemize}

Higher values across these metrics indicate that the compounded volume $x$ effectively integrates complementary content from the input views.

\subsubsection{Expert Evaluation}
In addition to quantitative metrics, qualitative image quality was systematically evaluated using a standardized expert-defined scoring scheme. It was assessed using a simple three-point ordinal scale. A score of 3 indicated good image quality, defined by clear delineation of anatomical structures and optimal visualization. A score of 2 indicated fair image quality, with acceptable but suboptimal structural definition. A score of 1 indicated poor image quality, characterized by limited anatomical detail and reduced diagnostic usefulness. The evaluation focused on the definition of key neuroanatomical structures, including visualization of the cerebral hemispheres, ventricles, basal ganglia, and posterior fossa. In addition, image quality was assessed based on the presence of motion-related artifacts and the reduction of image distortion, particularly improved slice interposition and continuity across adjacent slices. For each subject, reconstructions were compared within the same individual; the reconstruction with the best overall image quality was selected as the reference and assigned the highest score, and the remaining reconstructions were scored relative to that reference.

\begin{table*}[t]
\centering
\caption{Comparison of Ultrasound Compounding Strategies}
\label{tab:comparison_fusion_strategies}
\renewcommand{\arraystretch}{1.2}
\setlength{\tabcolsep}{6pt}
\resizebox{\textwidth}{!}{%
\begin{tabular}{p{1.7 cm} p{3.6cm} p{2.0cm} p{12.8cm}}
\toprule
\textbf{Category} & \textbf{Supervision} & \textbf{Computation} & \textbf{Key Remarks} \\
\midrule
Multi-scale & Unsupervised & Low & 
Captures structures at multiple resolutions; effective at enhancing edges and textures. \\

Transformation & Unsupervised & Low & 
Efficient; useful for global structure but may lose fine details. \\

Variational & Unsupervised / Self-supervised & Low–Moderate & 
Flexible optimization framework; integrates priors; requires solver design and tuning. \\

Deep Learning & Unsupervised / Self-supervised & Moderate–High & 
Highly flexible; capture complex patterns; requires data or priors (e.g., pretrained denoisers or network training). \\
\bottomrule
\end{tabular}%
}
\end{table*}

\subsection{Experiments}

\subsubsection{Dataset}
The fetal brain imaging data used in this work were acquired at Boston Children's Hospital (BCH). All participants provided written informed consent. The imaging studies were approved by BCH's Institutional Review Board and followed HIPAA guidelines. Data from a total of 15 fetuses were included. All fetuses underwent same-day ultrasound and MRI brain exams. The ultrasound images were acquired with a GE HealthCare Voluson Expert 22 system with a RM7C XDclear Wide Band Convex Volume 3D transducer (GE HealthCare Technologies, Chicago, IL). The ultrasound scan for each fetus included between 2 and 8 independent views. These volumes were acquired during routine clinical exam and reflected typical ultrasound image quality with common imperfections such as acoustic shadowing, speckle noise, and restricted fields of view. The MRI T2-weighted stacks-of-slices were used to reconstruct a super-resolved MRI volume using an existing algorithm \cite{xu2023nesvor}. The ultrasound views for each fetus were then individually registered to the corresponding MRI volume using a method that we have recently developed and rigorously validated \cite{zeng2025towards}. These pre-aligned views were then used as input to each of the compounding methods studied in this work.

For each view, binary brain masks were manually created and applied to the registered ultrasound volumes to extract the brain region and remove the surrounding background before compounding. These pre-aligned and masked ultrasound views were then used as input to all compounding methods evaluated in this study.

\subsubsection{Settings}

We evaluate ultrasound compounding strategies across four methodological categories: multi-scale, transformation-based, variational, and DL–based approaches. 
\paragraph*{(1) Multi-scale methods}  
From the multi-scale family, DoG stacks are constructed with $\sigma_{\text{scales}} = \{0.5, 1.0, 2.0, 4.0\}$, spanning from \emph{fine} to \emph{coarse} scales and yielding $L = 3$ bands. At each scale, the DoG maps are fused between views using the mean operator, and the resulting per-scale maps are subsequently integrated over scales by summation. A global band gain of $3.0$ is applied to the fused detail before adding it to the mean of all views.

\paragraph*{(2) Transformation-based methods}  
From the transformation-based family, \textit{PCA} fusion performs dimensionality reduction across stacked multi-view ultrasound volumes, decomposing voxel-wise intensity variations into orthogonal components that capture shared anatomical structures while suppressing noise and view-dependent artifacts. The fused representation is reconstructed as a weighted combination of the leading components, where weights are proportional to their explained variance ratios--representing the proportion of total variance each principal component accounts for. The number of components is automatically selected to retain at least 95\% of the cumulative explained variance.

\paragraph*{(3) Variational methods}  
Within the variational family, the \textit{Variational} method formulates compounding as an energy minimization problem combining three terms: an $\ell_1$ voxel-wise fidelity loss between the compounded output and input views, a gradient consistency term ($\beta = 200$) to preserve fine structural details, and total variation (TV) regularization ($\alpha = 1.0$) to suppress noise and promote spatial smoothness. Optimization is performed using the AdamW solver~\cite{kingma2014adam} with a learning rate of $0.01$ over $100$ iterations.

\paragraph*{(4) Deep learning–based methods}  
From the deep learning–based category, we evaluate two representative approaches: a \textit{scan-specific self-supervised learning} model that leverages inter-view consistency between the input views and the compounded output, and a \textit{PnP} framework that iteratively refines a coarse average using pretrained denoisers in a training-free manner.  

The scan-specific self-supervised learning framework was based on a 3D U-Net architecture with four encoding–decoding levels and shared encoder weights across all input views. Feature fusion across views was performed using Residual 3D Fusion Blocks, which combine per-view experts through a learnable gating network, with an optional self-attention module in the bottleneck to enhance global context modeling. A DoG–based skip enhancement was applied to emphasize salient structural details, with the fused DoG added to the network output for detail boosting. The network employed 64 base filters in the first convolutional layer. Training was performed using the AdamW optimizer~\cite{kingma2014adam} with a learning rate of $10^{-3}$ for up to 100 epochs, a batch size of 6, and automatic mixed precision (bfloat16)~\cite{micikevicius2017mixed}. Input volumes were randomly cropped into patches of size $72 \times 72 \times 72$ voxels during training, and processed in a tiling scheme with the same patch size and $48$-voxel overlap during inference. The training objective includes the $\ell_1$-norm and SSIM losses between the predicted and target compounded volumes.

For the PnP framework, we adapted the method from~\cite{zhang2021plug}, following the default configuration provided in the authors’ official repository and the procedure outlined in Algorithm~\ref{alg:pnp-fusion}. The algorithm was executed for $10$ iterations using the pretrained Gaussian denoiser DRUNet~\cite{zhang2021plug} as the prior. Volumes were processed slice-wise along the axial, coronal, and sagittal axes, and the three directional reconstructions were subsequently averaged to produce the final compounded volume. No task-specific training was performed.

\paragraph*{Implementation details}  
All methods were implemented in a patch-based manner, where volumetric patches of size $96 \times 96 \times 96$ voxels were extracted with a 48-voxel overlap along each dimension, unless otherwise specified. The compounded patches were blended by averaging across overlapping regions to ensure smooth and seamless transitions at patch boundaries. All experiments were performed on a laptop equipped with an Intel Core i9-14900HX CPU, 32 GB of system memory, and an NVIDIA GeForce RTX 4090 Laptop GPU with 16 GB of VRAM, running Ubuntu 22.04 LTS.

\section{Results}
\begin{figure*}[h]
    \centering
    \includegraphics[width=\linewidth]{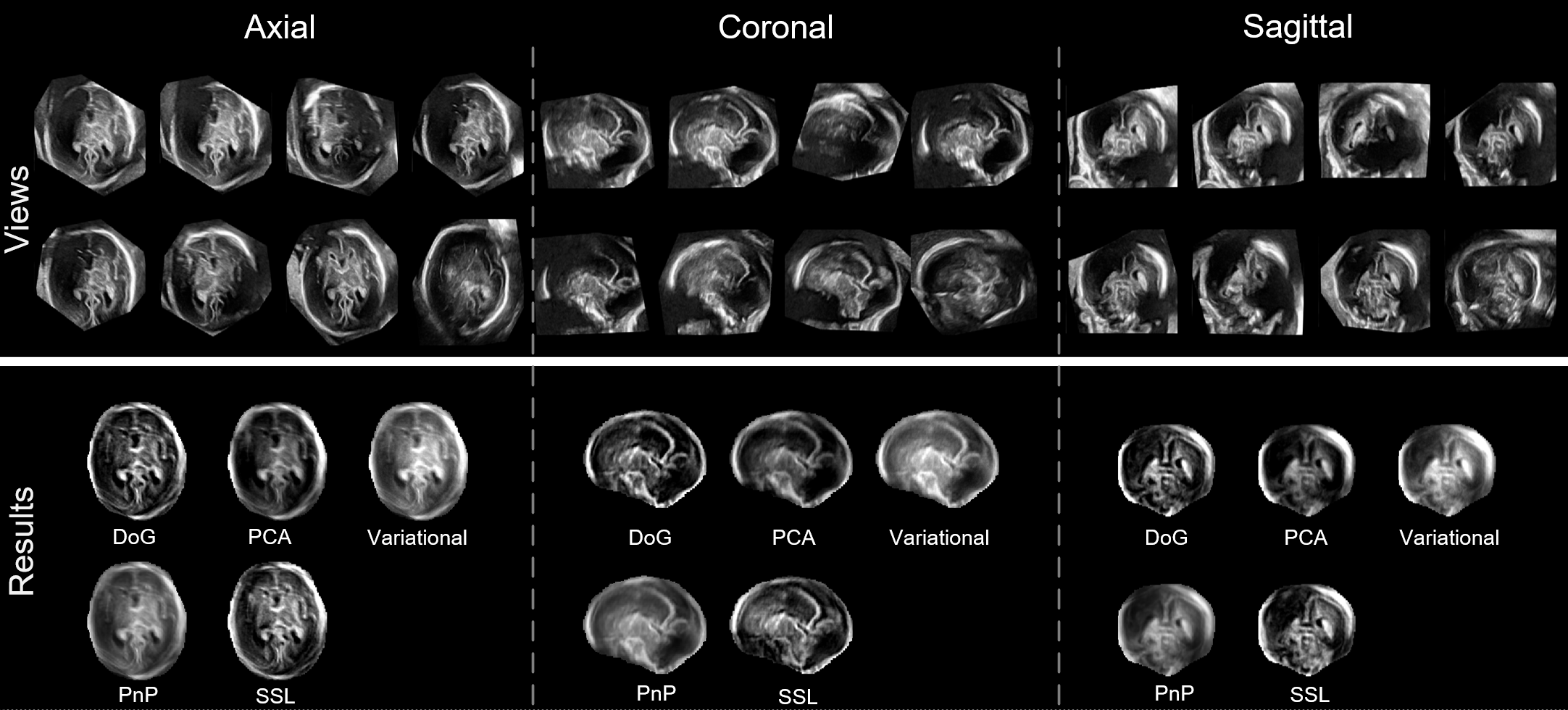}
    \caption{
    Compounding results from five representative methods applied to a fetal brain ultrasound dataset acquired from a single subject (Subject 8, eight 3D input views). The upper panel shows the input multi-view ultrasound volumes, and the lower panel displays the compounded outputs generated by the DoG, PCA, Variational, PnP, and SSL methods. Qualitative expert ratings (1–3 scale; higher indicates better perceptual quality) for this subject were: DoG = 3, PCA = 1, Variational = 2, PnP = 1, and SSL = 2.
    }
    \label{fig:results_subject8}
\end{figure*}

\begin{figure*}[h]
    \centering
    \includegraphics[width=\linewidth]{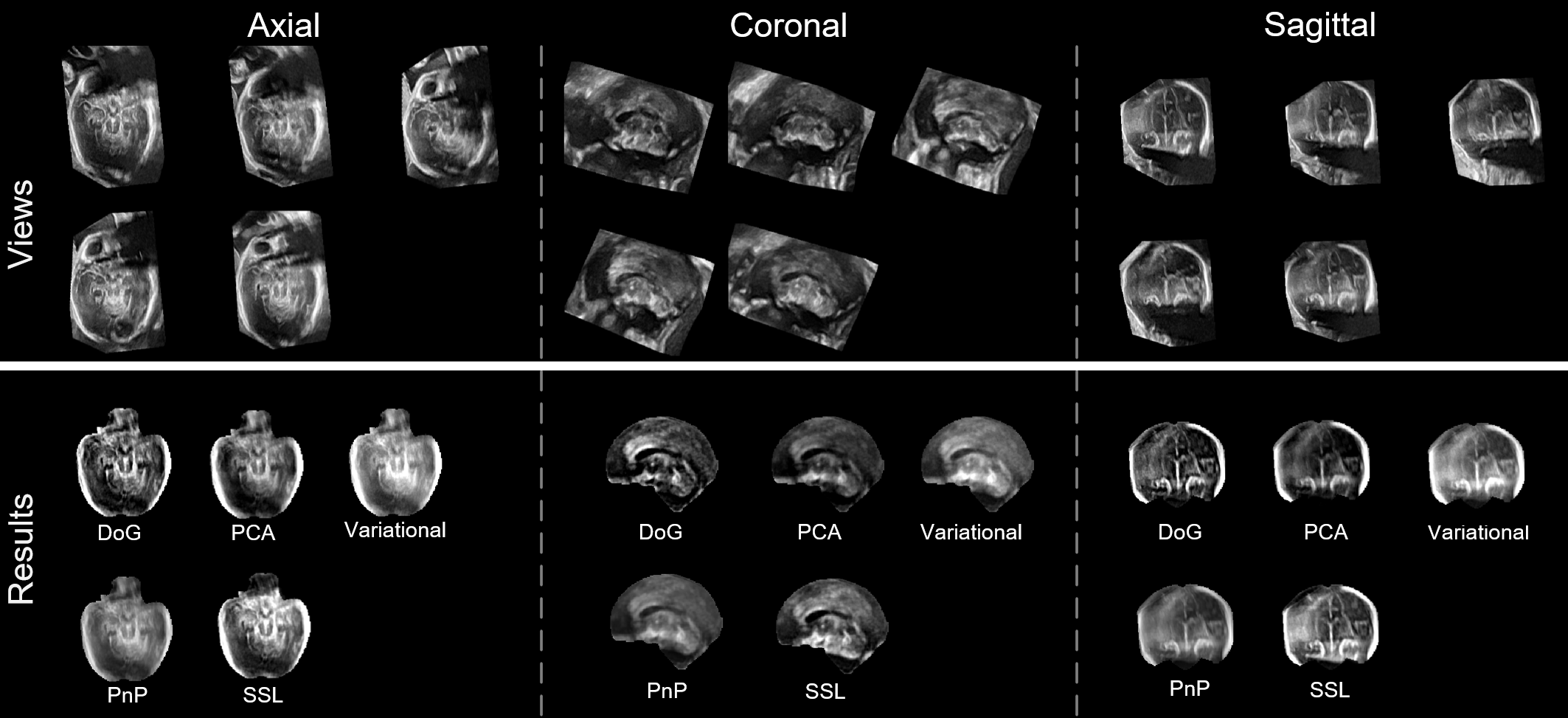}
    \caption{
    Compounding results from five representative methods applied to a fetal brain ultrasound dataset acquired from a single subject (Subject 5, five 3D input views). The upper panel shows the input multi-view ultrasound volumes, and the lower panel displays the compounded outputs generated by the DoG, PCA, Variational, PnP, and SSL methods. Qualitative expert ratings (1–3 scale; higher indicates better perceptual quality) for this subject were: DoG = 3, PCA = 2, Variational = 1, PnP = 2, and SSL = 2.
    }
    \label{fig:results_subject5}
\end{figure*}

The compounding results from representative methods across all categories—DoG (multi-scale), PCA (transformation-based), Variational, and deep learning–based approaches, including PnP (training-free) and SSL (self-supervised learning)—were evaluated on ten multi-view fetal ultrasound datasets. Quantitative performance, assessed using well-established image quality metrics widely adopted in the image fusion literature, together with qualitative scores provided by an expert radiologist, is summarized in Table~\ref{tab:compounding_results}. Detailed results for individual samples are provided in the supplementary Table~\ref{tab:UScompounding_restults_all}. For brevity, qualitative comparisons for two representative samples are presented in Figs.~\ref{fig:results_subject8}–\ref{fig:results_subject5}.

\begin{table*}[ht]
\centering
\caption{Quantitative comparison of compounding methods across all samples (mean $\pm$ standard deviation). 
Qualitative scores are based on expert radiologist ratings (1--3 scale) and were rounded to one decimal place to reflect the ordinal nature of the rating scale.}
\label{tab:compounding_results}
\begin{tabular}{lccccc|c}
\toprule
\textbf{Method} & \textbf{SSIM} & \textbf{PSNR} & \textbf{CC} & \textbf{Entropy} & \textbf{MI} & \textbf{Qualitative Score} \\
\midrule
DoG         & $0.779 \pm 0.042$ & $20.9 \pm 1.54$ & $0.858 \pm 0.045$ & $0.631 \pm 0.088$ & $2.25 \pm 0.353$ & $3.0 \pm 0.0$ \\
PCA         & $0.795 \pm 0.041$ & $21.4 \pm 1.81$ & $0.883 \pm 0.050$ & $0.658 \pm 0.081$ & $2.31 \pm 0.323$ & $1.9 \pm 0.3$ \\
Variational & $0.813 \pm 0.051$ & $19.9 \pm 1.84$ & $0.916 \pm 0.047$ & $0.793 \pm 0.100$ & $2.63 \pm 0.350$ & $1.7 \pm 0.5$ \\
PnP         & $0.592 \pm 0.064$ & $16.3 \pm 1.38$ & $0.916 \pm 0.048$ & $0.765 \pm 0.100$ & $2.57 \pm 0.338$ & $1.5 \pm 0.5$ \\
SSL         & $0.799 \pm 0.054$ & $20.8 \pm 1.94$ & $0.880 \pm 0.068$ & $0.733 \pm 0.093$ & $2.54 \pm 0.332$ & $1.8 \pm 0.4$ \\
\bottomrule
\end{tabular}
\end{table*}

Fig.~\ref{fig:results_subject8} illustrates the compounding results for a representative sample with eight 3D views (subject 8), where each view was masked to exclude background and non-brain regions. The outputs across different methods exhibit distinct visual characteristics. The DoG-based approach effectively emphasizes multi-scale salient features while successfully suppressing noise-like artifacts, achieving a good balance between structural preservation and detail enhancement. PCA-based fusion focuses on energy maximization and preserves overall variance across views, though some fine anatomical details are lost due to its transformation nature. The Variational method struggles to maintain consistent salient structures across views, resulting in over-smoothed outputs. The PnP approach preserves major anatomical regions but tends to smooth out finer structural details. In contrast, the SSL-based method demonstrates superior preservation of salient anatomical features while maintaining cross-view smoothness and coherence.

For quantitative comparison in the case of Subject 8 (Table~\ref{tab:UScompounding_restults_all}), the Variational method achieved one of the highest correlation coefficients (CC = 0.842) and mutual information (MI = 0.848), indicating strong statistical consistency across views. Similarly, the PnP approach reported a high CC (0.841) but a comparatively low PSNR (14.66), suggesting limited reconstruction fidelity despite numerical correlation. In contrast, the DoG method achieved lower quantitative scores (SSIM = 0.717, PSNR = 18.45) while producing visually sharper and more distinct structural boundaries. The SSL approach (SSIM = 0.711, PSNR = 17.95) preserved local feature continuity and salient anatomical patterns, despite achieving only moderate quantitative metric values. Comparing these observations with the qualitative scores assigned by the expert radiologist illustrates that higher quantitative metric values do not necessarily correspond to improved perceptual or diagnostic image quality in fetal ultrasound compounding.

Fig.~\ref{fig:results_subject5} presents the compounding results from five representative methods applied to a fetal brain ultrasound dataset acquired from a single subject with five 3D input views (subject 5). The upper panel shows the input multi-view volumes after masking to remove background and non-brain regions, while the lower panel displays the corresponding compounded results. The DoG method effectively enhances edge contrast and preserves fine structural details, particularly around cortical boundaries. The PCA method maintains overall smoothness and preserves dominant intensity structures but tends to blur subtle internal features. The Variational approach produces over-smoothed results, reducing local contrast and anatomical sharpness. The PnP method reconstructs major regions but suppresses high-frequency details, leading to slightly flattened textures. In contrast, the SSL-based compounding achieves balanced sharpness and smoothness, preserving salient anatomical features with higher inter-view consistency compared to the other approaches.

Although the numerical results in Table~\ref{tab:UScompounding_restults_all} indicate high similarity scores across several methods for Subject 5, these values do not always correspond to perceptually superior or diagnostically meaningful outcomes as assessed by the expert radiologist. For example, the Variational and PnP methods exhibit high correlation coefficients and mutual information, indicating strong statistical consistency; however, their outputs in Fig.~\ref{fig:results_subject5} appear over-smoothed and lack fine structural detail. In contrast, methods such as DoG and SSL achieve more moderate yet balanced quantitative scores (e.g., SSIM and PSNR) while producing visually sharper and more anatomically coherent reconstructions. Overall, this observation further highlights the limitations of conventional image quality metrics in ultrasound compounding, where perceptual and clinical relevance may not fully align with purely statistical measures.

Finally, we comment on the limitations of quantitative metrics. The observed discrepancy between quantitative image quality metrics and expert radiologist assessments highlights a fundamental limitation of conventional evaluation measures in fetal ultrasound compounding. Widely used metrics such as SSIM, PSNR, correlation coefficient, and mutual information primarily capture statistical similarity and noise suppression, but are often insensitive to clinically relevant attributes, including fine anatomical detail and structural clarity. As a result, excessive smoothing or intensity homogenization may artificially inflate metric values while degrading perceptual quality and diagnostic interpretability.

\section{Conclusion}

This study provides the first systematic categorization of computational strategies for fetal brain ultrasound compounding, encompassing both classical and deep learning–based frameworks. We implemented representative methods from four major categories—multi-scale (DoG), transformation-based (PCA), variational, and deep learning approaches (PnP and SSL)—and rigorously evaluated them using standard quantitative image-quality metrics, supplemented by expert radiologist assessments.

Experiments on ten multi-view fetal brain datasets showed that, while conventional quantitative metrics capture statistical similarity, they do not consistently reflect perceptual or diagnostic image quality. Consequently, the expert visual assessment remains essential for meaningful evaluation of ultrasound compounding methods. The results also revealed that each approach offers distinct strengths: DoG enhances salient anatomical features while suppressing noise effectively; PCA provides stable fusion through statistical decorrelation; variational and PnP methods offer flexible regularization; and SSL yields task-adaptive representations without requiring explicit supervision.

All implementations are publicly released as the open-source USFetal Compounding Toolbox, a computationally efficient and lightweight platform that runs smoothly on GPU-equipped laptops, enabling practical deployment in both research environments and clinical centers.

Looking ahead, future research should focus on unified frameworks that perform registration and compounding simultaneously to improve anatomical consistency under fetal motion and view-dependent artifacts. Moreover, integrating low-level vision tasks (alignment and compounding) with high-level clinical objectives (segmentation, anomaly detection, etc) can enable task-aware compounding that is diagnostically meaningful to real-world clinical workflows. Taken together, our experiments reveal that conventional quantitative metrics often fail to reflect clinically meaningful image quality, highlighting the need for evaluation criteria that better align with expert perception and clinical relevance.

\section{Acknowledgment}
The authors would like to thank the Research Computing Center at Boston Children's Hospital and Harvard Medical School, USA, as well as the CSC–IT Center for Science, Finland, for their support in providing computational resources.

\section{Supplementary Materials}

The quantitative performance measures across all subjects are presented in Table~\ref{tab:UScompounding_restults_all}, providing a comprehensive comparison among the evaluated methods.

\begin{table*}[ht]
\centering
\caption{Quantitative and qualitative comparison of compounding methods across individual subjects. 
Quantitative metrics are reported per subject, and qualitative scores represent raw expert radiologist ratings on a 1--3 ordinal scale.}
\label{tab:UScompounding_restults_all}
\resizebox{0.56\textwidth}{0.34\textheight}{%
\begin{tabular}{clccccc|c}
\toprule
\textbf{Subject ID} & \textbf{Method} & \textbf{SSIM} & \textbf{PSNR} & \textbf{CC} & \textbf{Entropy} & \textbf{MI} & \textbf{Qualitative Score} \\
\toprule

\multirow{6}{*}{\rotatebox[origin=c]{90}{\textbf{Subject 1}}}
 & DoG         & 0.7333 & 19.29 & 0.7959 & 0.5094 & 1.978  & 3 \\
 & PCA         & 0.7483 & 19.00 & 0.8253 & 0.5709 & 2.190  & 2\\
 & Variational & 0.7503 & 17.88 & 0.8466 & 0.6949 & 2.498  & 2\\
 & PnP         & 0.6268 & 18.04 & 0.8459 & 0.6552 & 2.415  & 2\\
 & SSL         & 0.7269 & 18.66 & 0.7739 & 0.5737 & 2.342  & 1\\
\midrule

\multirow{6}{*}{\rotatebox[origin=c]{90}{\textbf{Subject 2}}}
 & DoG         & 0.7792 & 20.58 & 0.8682 & 0.5896 & 2.182 & 3\\
 & PCA         & 0.8103 & 20.59 & 0.9061 & 0.6374 & 2.306 & 2\\
 & Variational & 0.8078 & 18.43 & 0.9304 & 0.7427 & 2.537 & 2\\
 & PnP         & 0.6139 & 18.65 & 0.9335 & 0.6913 & 2.422 & 2\\
 & SSL         & 0.7905 & 18.95 & 0.9099 & 0.7102 & 2.505 & 2\\
\midrule

\multirow{6}{*}{\rotatebox[origin=c]{90}{\textbf{Subject 3}}}
 & DoG         & 0.7755 & 22.16 & 0.8943 & 0.6305 & 2.552 & 3\\
 & PCA         & 0.7904 & 22.82 & 0.9292 & 0.6497 & 2.570 & 2\\
 & Variational & 0.8307 & 21.11 & 0.9577 & 0.8453 & 3.071 & 2\\
 & PnP         & 0.5456 & 16.39 & 0.9605 & 0.8145 & 2.960 & 1\\
 & SSL         & 0.8185 & 22.04 & 0.9430 & 0.7949 & 2.986 & 2\\
\midrule

\multirow{6}{*}{\rotatebox[origin=c]{90}{\textbf{Subject 4}}}
 & DoG         & 0.7601 & 19.82 & 0.8360 & 0.6541 & 2.063 & 3\\
 & PCA         & 0.7778 & 20.45 & 0.8693 & 0.7338 & 2.283 & 2\\
 & Variational & 0.7886 & 19.02 & 0.8888 & 0.8884 & 2.617 & 2\\
 & PnP         & 0.5644 & 16.01 & 0.8876 & 0.8614 & 2.579 & 1\\
 & SSL         & 0.7685 & 19.37 & 0.8484 & 0.7997 & 2.493 & 1\\
\midrule

\multirow{6}{*}{\rotatebox[origin=c]{90}{\textbf{Subject 5}}}
 & DoG         & 0.8682 & 22.59 & 0.8884 & 0.5117 & 1.670 & 3\\
 & PCA         & 0.8715 & 22.86 & 0.8926 & 0.5043 & 1.644 & 2\\
 & Variational & 0.8789 & 21.25 & 0.9405 & 0.5663 & 1.776 & 1\\
 & PnP         & 0.7585 & 16.57 & 0.9413 & 0.5578 & 1.765 & 2\\
 & SSL         & 0.8730 & 21.78 & 0.9052 & 0.5621 & 1.785 & 2\\
\midrule

\multirow{6}{*}{\rotatebox[origin=c]{90}{\textbf{Subject 6}}}
 & DoG           & 0.7962 & 21.76 & 0.8944 & 0.6855 & 2.604 & 3\\
 & PCA           & 0.8139 & 22.61 & 0.9164 & 0.6894 & 2.590 & 2\\
 & Variational   & 0.8430 & 21.72 & 0.9602 & 0.8090 & 2.879 & 1\\
 & PnP           & 0.5726 & 14.41 & 0.9599 & 0.7893 & 2.843 & 1\\
 & SSL           & 0.8172 & 22.71 & 0.9280 & 0.7737 & 2.785 & 2\\
\midrule

\multirow{6}{*}{\rotatebox[origin=c]{90}{\textbf{Subject 7}}}
 & DoG          & 0.7842 & 20.61 & 0.8650 & 0.7496 & 2.456 & 3\\
 & PCA          & 0.8000 & 21.28 & 0.8896 & 0.7578 & 2.471 & 2\\
 & Variational  & 0.8132 & 20.21 & 0.9262 & 0.8171 & 2.591 & 2\\
 & PnP          & 0.5684 & 14.88 & 0.9231 & 0.8042 & 2.579 & 2\\
 & SSL          & 0.8090 & 21.03 & 0.8976 & 0.8267 & 2.639 & 2\\
\midrule

\multirow{6}{*}{\rotatebox[origin=c]{90}{\textbf{Subject 8}}}
 & DoG            & 0.7173 & 18.45 & 0.7878 & 0.7659 & 2.732 & 3\\
 & PCA            & 0.7207 & 18.62 & 0.7903 & 0.7493 & 2.690 & 1\\
 & Variational    & 0.7262 & 17.10 & 0.8422 & 0.8480 & 2.889 & 2\\
 & PnP            & 0.5569 & 14.66 & 0.8405 & 0.8247 & 2.851 & 1\\
 & SSL            & 0.7110 & 17.95 & 0.7648 & 0.7851 & 2.785 & 2\\
\midrule

\multirow{6}{*}{\rotatebox[origin=c]{90}{\textbf{Subject 9}}}
 & DoG          & 0.8110 & 23.39 & 0.9257 & 0.5719 & 1.847 & 3\\
 & PCA          & 0.8234 & 24.33 & 0.9577 & 0.6146 & 1.939 & 2\\
 & Variational  & 0.8896 & 22.79 & 0.9776 & 0.9036 & 2.645 & 2\\
 & PnP          & 0.5528 & 16.63 & 0.9785 & 0.8699 & 2.593 & 2\\
 & SSL          & 0.8749 & 23.84 & 0.9652 & 0.7517 & 2.377 & 2\\
\midrule

\multirow{6}{*}{\rotatebox[origin=c]{90}{\textbf{Subject 10}}}
 & DoG         & 0.7625 & 20.52 & 0.8292 & 0.6377 & 2.371 & 3\\
 & PCA         & 0.7891 & 21.65 & 0.8559 & 0.6708 & 2.467 & 2\\
 & Variational & 0.8000 & 19.09 & 0.8929 & 0.8186 & 2.797 & 1\\
 & PnP         & 0.5589 & 16.69 & 0.8896 & 0.7818 & 2.734 & 1\\
 & SSL         & 0.8001 & 21.40 & 0.8621 & 0.7567 & 2.695 & 2\\
\midrule

\end{tabular}}
\end{table*}

\bibliographystyle{ieeetr}
\bibliography{main.bib}

\end{document}